\documentclass[twocolumn,showpacs,superscriptaddress,preprintnumbers,amsmath,amssymb,epsfig,floatfix,aps,bibnotes]{revtex4-1}\usepackage{amssymb}
\usepackage{multirow}
\usepackage{lipsum}
\usepackage{amsmath}
\usepackage{upgreek}
\usepackage{epsfig}
\usepackage{graphicx}
\usepackage{dcolumn}
\usepackage{color}
\usepackage{natbib}  
\usepackage{hyperref}
\usepackage{breakurl}
\usepackage{mathrsfs}
\hypersetup{colorlinks=true, citecolor=blue, urlcolor=blue, linkcolor=blue}
\usepackage{bm}
\usepackage{tabularx}
\newcolumntype{L}[1]{>{\raggedright\arraybackslash}p{#1}}
\newcolumntype{C}[1]{>{\centering\arraybackslash}p{#1}}
\newcolumntype{R}[1]{>{\raggedleft\arraybackslash}p{#1}}

\usepackage[capitalize]{cleveref}
\definecolor{myblue}{RGB}{0, 50, 125}

\begin{document}
\title{Ferromagnetism in nitrogen doped graphene}
\author{Rohit Babar}
\affiliation{Department of Physics, Indian Institute of Science Education and Research, Pune 411008, India}	
\author{Mukul Kabir}
\email{mukul.kabir@iiserpune.ac.in} 
\affiliation{Department of Physics, Indian Institute of Science Education and Research, Pune 411008, India}
\affiliation{Centre for Energy Science, Indian Institute of Science Education and Research, Pune 411008, India}
\date{\today}

\begin{abstract} 
Inducing a robust long-range magnetic order in diamagnetic graphene remains a challenge. While nitrogen-doped graphene is reported to be a promising candidate, the corresponding exchange mechanism endures unclear and is essential to tune further and manipulate magnetism. Within the first-principles calculations, we systematically investigate the local moment formation and the concurrent interaction between various defect complexes. The importance of adatom diffusion on the differential defect abundance is discussed. The individual nitrogen complexes that contribute toward itinerant and a local magnetic moment are identified. The magnetic interaction between the complexes is found to depend on the concentration, complex type, sublattice, distance, and orientation. We propose that the direct exchange mechanism between the delocalized magnetic moment originating from the itinerant $\pi$-electron at the prevalent graphitic complexes to be responsible for the observed ferromagnetism. We show that B co-doping further improves ferromagnetism. Present results will assist in the microscopic understanding of the current experimental results and motivate experiments to produce robust magnetism following the proposed synthesis strategy.
\end{abstract}
\maketitle

\section{Introduction}
The absence of intrinsic magnetism in graphene limits its application in spintronics and is being actively persuaded since it was first isolated.~\citep{Novoselov666,C7CS00288B} However, the magnetism can be induced in graphene due to the $p_z$ electron imbalance in bipartite hexagonal sublattice,~\citep{PhysRevB.77.035427,PhysRevLett.109.186604, Herrero437} electron localization at the zigzag edge,~\citep{PhysRevB.54.17954,PhysRevLett.99.177204,PhysRevLett.100.047209,PhysRevB.90.035403,nature13831,PhysRevB.95.174419,s41586-018-0154-7} interaction with a spin-orbit coupled ferromagnetic substrate,~\citep{nn303771f,PhysRevLett.114.016603,2053-1583-4-1-014001} structural defect creation,~\citep{PhysRevLett.91.227201,PhysRevB.75.125408, nphys1399, PhysRevLett.104.096804,10.1038/nphys2183,ncomms3010,PhysRevLett.117.166801} and heteroatom doping.~\citep{PhysRevLett.102.126807, jacs.6b12934} However, a long-range magnetic order in graphene can only emerge through the exchange coupling between these induced local moments. 

Hydrogen absorption on graphene creates an imbalance in $p_z$ electrons between the two sublattices, which in turn generates a net magnetic moment.~\citep{Herrero437} Further, the spin-polarized state is observed to extend over several nanometers through the scanning tunneling microscopy (STM) experiment. This spatially extended magnetic state provides a channel for direct coupling between the induced magnetic moments at large distances. The half-filled flat band at the Fermi level produces localized magnetic moments at the zigzag edge of a graphene flake,~\citep{PhysRevB.54.17954} and a long-range order has been predicted.~\citep{PhysRevLett.99.177204,PhysRevLett.100.047209,PhysRevB.90.035403,nature13831,PhysRevB.95.174419,s41586-018-0154-7}  However,  the edge magnetism is predicted to be very intricate as it is strongly affected by the shape and size of the flakes, charge doping, chemical functionalization at the edge, and on-site Coulomb interaction.~\citep{PhysRevB.90.035403,PhysRevB.95.174419} Proximity-induced exchange-coupled ferromagnetic order sets in graphene sheet at room temperature while places on an atomically flat ferromagnetic thin film insulator.~\citep{nn303771f,PhysRevLett.114.016603,2053-1583-4-1-014001} The smallest structural defect, a lattice vacancy in graphene generates a semi-localized magnetic moment due to the $\pi$ electron imbalance in addition to a $\sigma$ contribution from the dangling bond.~\citep{PhysRevB.75.125408,PhysRevLett.104.096804,10.1038/nphys2183,ncomms3010,PhysRevLett.117.166801}  Thus, proton irradiated graphene and graphene with a network of point defects show magnetic ordering.~\citep{PhysRevLett.91.227201,nphys1399} Similarly, the larger lattice defects such as grain boundaries with dangling bonds are theoretically predicted to have magnetic moment.~\citep{PhysRevB.85.115407} However, the experimental evidence of a magnetic ordering remains scarce and controversial.~\citep{nphys1399,10.1038/nphys2183,nl802810g,jp903397u,PhysRevLett.105.207205,10.1063/1.3628245} While a room temperature ferromagnetism is predicted,~\citep{nphys1399,nl802810g,jp903397u} defected graphene with vacancies and other $sp^3$ defects lead to paramagnetism and devoid of any magnetic ordering down to low temperature.~\citep{10.1038/nphys2183,PhysRevLett.105.207205,10.1063/1.3628245} Furthermore, in the cases where ferromagnetic signals have been detected, the origin of such long-range ordering remained unexplained. The presence of $d$-electrons in transition-metal impurities induces a local moment in pristine and defected graphene.~\citep{PhysRevLett.102.126807,PhysRevB.93.045433} However, only a paramagnetic ordering has been observed in this transition-metal doped graphene samples that were investigated through the X-ray magnetic circular dichroism measurement.~\citep{PhysRevLett.110.136804} 

In contrast, the substitutional and $sp^3$ impurities have attracted much attention with intriguing but elusive magnetic properties. 
The itinerant $sp$-electron Stoner ferromagnetism in a narrow impurity band is fundamentally different from the magnetism induced by the $d$-band of transition metals.~\citep{0953-8984-18-31-016} Such $sp$-electron ferromagnetism may lead to a higher Curie temperature than the usual dilute magnetic semiconductors. The adatom impurities like hydrogen and fluorine exhibit strong absorption and induce magnetic moments~\citep{PhysRevB.77.035427,PhysRevLett.109.186604,PhysRevB.83.085410,PhysRevB.87.174435} due to sublattice imbalance in $\pi$-electrons following Lieb's theorem.~\citep{PhysRevLett.62.1201} While the hydrogenated graphene show magnetic order at 5 K,~\citep{Herrero437} the fluorine doped graphene reveals paramagnetic behaviour.~\citep{nphys1399,ncomms3010} In this context, the substitutional N-doped graphene attracted much attention, which exhibits robust ferromagnetism.~\citep{ja512897m,jacs.6b12934,Liu2016,LI2015460,Miao2016,C3NR34291C} However, the corresponding Curie temperature and saturation magnetization are reported to show a strong dependence on the growth, annealing strategies, and nitrogen concentration, which hinder a careful microscopic analysis. A variety of nitrogen complexes are generated during the non-equilibrium synthesis, while N-concentration and the type of defect complex determines the electronic and transport properties.~\citep{jacs.6b12934,ja512897m,jacs.6b12934,Liu2016,LI2015460,Miao2016,C3NR34291C,Zhao999,nl2031037,cs200652y,srep00586,ja408463g,10.1021/nn506074u,PhysRevB.84.245446,nl304351z,cm102666r} Recently, the room-temperature ferromagnetism was reported in boron and nitrogen-doped turbostratic carbon films grown via pulsed laser deposition technique, while the saturation magnetization is observed to increase with increasing N-doping.~\citep{unpublished} However, even with such efforts, the nature and microscopic mechanism of magnetic interaction are not well understood in N-doped graphene, which is necessary to tune and manipulate magnetism toward its successful applications. 


Within the first-principles calculations, here we systematically investigate the electronic and magnetic properties of various N-defect complexes in graphene. Further, how these defects interact with each other and lattice vacancies are studied to understand the mechanism responsible for the observed ferromagnetism. We report that the long-range magnetic interactions show a strong dependence on the N-concentration, complex type and on which sublattice they reside, distance, and orientation. We observe that the magnetism in the graphitic and vacancy containing N-complexes are fundamentally very different, and propose that the observed ferromagnetic order originates from the direct exchange between the extended magnetic states of the graphitic defects. Further, we investigate the crucial role of boron co-doping in the ferromagnetism.

\section{Computational details}
The spin-polarized density functional theory based calculations~\citep{PhysRevB.47.558,PhysRevB.54.11169} were performed within the projector augmented wave formalism,~\citep{PhysRevB.50.17953}, and the wave functions were expanded in a plane-wave basis with 600 eV kinetic energy cut-off. The exchange-correlation energy was described using the conventional Perdew-Burke-Ernzerhof (PBE) form for the generalized gradient approximation (GGA).~\citep{PhysRevLett.77.3865} The structures were fully relaxed until all the forces became smaller than 0.015 eV/\AA~ threshold using the PBE functional. The Brillouin zone was sampled using a $\Gamma$-centered 8$\times$6$\times$1 $k$-grid for the structural optimization, and a much dense $\Gamma$-centered 24$\times$18$\times$1 $k$-grid was used for the evaluation of density of states (DOS) and magnetization density. The subsequent N-doping was carried with the 4$\times$3$(\sqrt3,3)a_0$ and 10$\times$5$(\sqrt3,3)a_0$ rectangular supercells, where $a_0$ is the C--C bond length in graphene.  The periodic images of the supercell along the direction perpendicular to the graphene sheet are separated by at least 12 \AA\ distance. It is known that the self-interaction error inherent to the conventional GGA functional results in spurious electron delocalization and incorrect magnetic solutions for adatom doping and vacancy defects in graphene.~\citep{PhysRevB.87.174435,10.1021/acs.jpcc.7b02306,PhysRevB.96.125431}  In this regard, we recalculated the electronic and magnetic properties using the Heyd-Scuseria-Ernzerhof (HSE06)~\citep{10.1063/1.1564060} hybrid functional as well as using the strongly constrained and appropriately normed (SCAN) meta-GGA functional.~\citep{PhysRevLett.115.036402} While the inclusion of fractional exact exchange for HSE06 hybrid functional improves the description of the electronic structure and the magnetization density, it is computationally expensive. In contrast, since the semilocal SCAN meta-GGA density functional satisfies all the known exact constraints for the exchange-correlation energy and the self-interaction error is reduced, it should better describe the thermodynamic, electronic and magnetic properties of materials.  Indeed the SCAN functional predicts these properties correctly for diverse materials with a similar computational cost to the conventional GGA.~\citep{Sun2016,PhysRevMaterials.2.063801} In the present context of defect magnetism and the concurrent interaction between them, we further validate the applicability of the SCAN functional. The results are obtained within the SCAN functional if it is not otherwise stated. 

\begin{figure}[!t]
 \begin{center}
\rotatebox{0}{\includegraphics[width=0.48\textwidth]{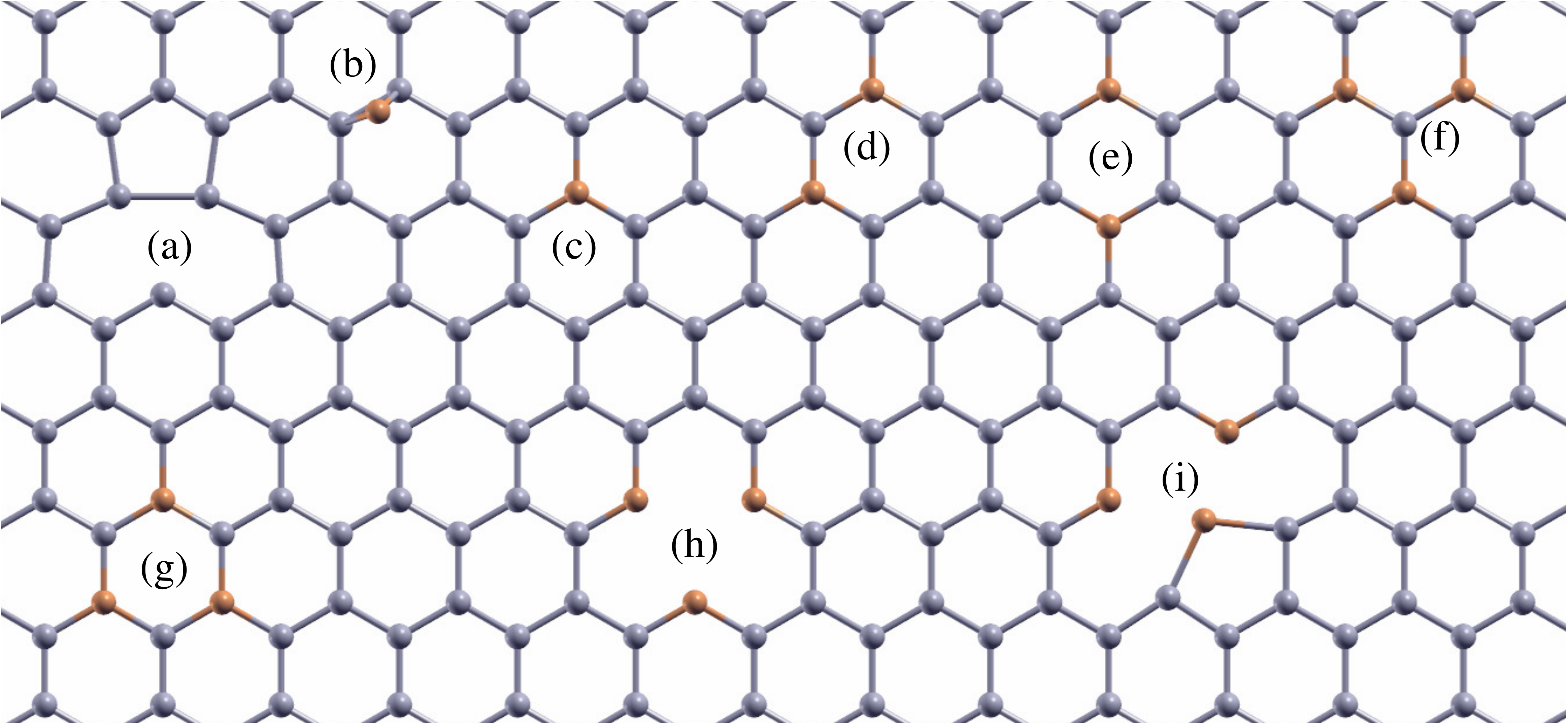}}
 \caption{Schematic representation of defects in N-doped graphene. Apart from the lattice vacancies (a), chemisorbed (b), various graphitic [single graphitic N (c), meta-graphitic N pair N$_{\rm AA}$ (d), para-graphitic N pair N$_{\rm AB'}$ (e), graphitic 3N (f), triazine (g)], and different vacancy containing N-complexes [pyridinic (h), pyrrolic (i)] are investigated.  A and B represent the two graphene sublattice.}
 \label{fig:figure1}
 \end{center}
 \end{figure}

\section{Results and Discussion}
Before we investigate the magnetic interactions between the moment generating defect complexes in N-doped graphene, we first address the local magnetism produced by these individual defects. In addition to the lattice vacancies, the common defects in N-doped graphene are chemisorbed, graphitic and vacancy containing nitrogen complexes  (Figure~\ref{fig:figure1}).~\citep{jacs.6b12934,ja512897m,jacs.6b12934,Liu2016,LI2015460,Miao2016,C3NR34291C,Zhao999,PhysRevB.84.245446,nl2031037,cs200652y,srep00586,nl304351z,ja408463g,10.1021/nn506074u} As the treatment of exchange-correlation may affect the electronic and magnetic description of these defects, here we compare the PBE, hybrid HSE06, and SCAN meta-GGA functionals. 

\subsection{Magnetism of individual defects}
\begin{figure}[!t]
 \begin{center}
\rotatebox{0}{\includegraphics[width=0.48\textwidth]{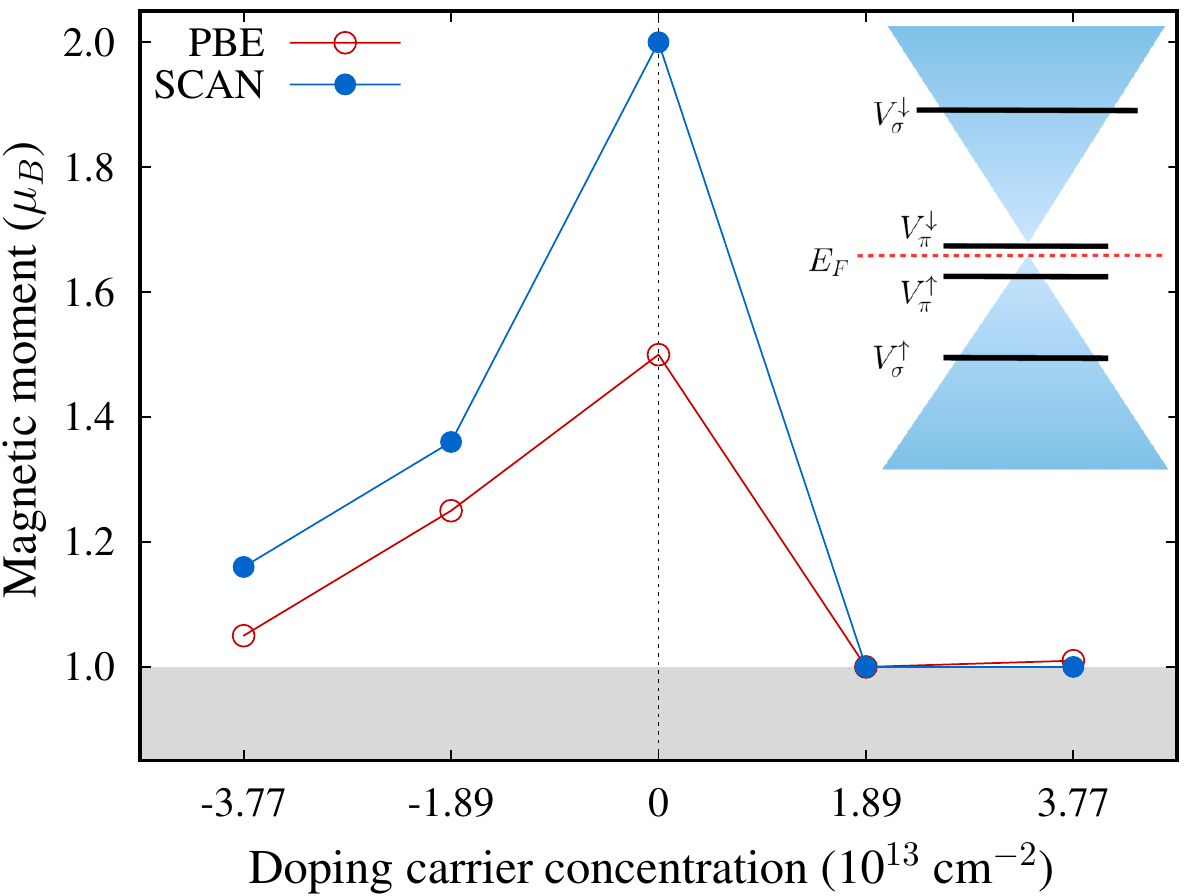}}
 \caption{The dual origin of vacancy magnetism in graphene and its evolution with carrier doping. Positive (negative) concentration indicates electron (hole) doping. While the $\pi$-component of the magnetic moment is quenched under carrier doping, the $\sigma$-component remains protected. The inset represents a schematic energy diagram indicating defect induced and spin-split localized $\pi$ and $\sigma$ states along with the bulk $\pi$ bands for the Jahn-Teller distorted V$_1(5|9)$ vacancy. This schematic picture explains the evolution of the magnetic moment of a lattice vacancy with varied carrier doping. While the magnetism is correctly described by the SCAN meta-GGA exchange-correlation functional, the PBE-GGA functional underestimates the moment.}
 \label{fig:figure2}
 \end{center}
 \end{figure}

{\em Single vacancy.} Magnetic properties of single vacancy defect in graphene are still studied and debated.~\cite{PhysRevB.75.125408,10.1038/nphys2183,ncomms3010,10.1021/acs.jpcc.7b02306,PhysRevB.96.125431} The Jahn-Teller distorted V$_1(5|9)$ vacancy [Figure~\ref{fig:figure1}(a)] generates defect induced localized $\sigma$ and $\pi$ states, V$_{\sigma}$ and V$_{\pi}$ (Figure~\ref{fig:figure2}). Thus, the magnetic moment of V$_1(5|9)$ vacancy has two components.~\citep{PhysRevLett.117.166801,ncomms3010} The $\pi$-electron imbalance in the bipartite graphene lattice should generate 1 $\mu_B$ moment according to the Lieb's theorem. Additionally, the singly occupied dangling $\sigma$-bond should provide with another 1 $\mu_B$ moment, leaving the total vacancy moment to 2 $\mu_B$. However, the $\pi$-magnetism is suppressed within the conventional GGA description of exchange-correlation due to improper delocalization of electrons in the V$_{\pi}$ orbital. Thus, the magnetic moment of V$_1(5|9)$ is underestimated to 1.5 $\mu_B$ within the PBE exchange-correlation functional (Figure~\ref{fig:figure2}, Table~\ref{tab:table1}), similar to that predicted earlier.~\cite{PhysRevB.75.125408} Further, the moment depends on the supercell size, and thus on the vacancy concentration.~\citep{supple} The picture is slightly improved while a fraction $a_{_{\rm  HF}}$ of Hartree-Fock exact exchange is considered in the HSE06 functional and 1.6 $\mu_B$ moment is predicted at the vacancy for $a_{_{\rm HF}}$=0.25. A further increase in $a_{_{\rm  HF}}$ to 0.4 ensures a 2 $\mu_B$ moment at the vacancy. In contrast, treating the exchange-correlation with SCAN meta-GGA functional describe the magnetism correctly with 2 $\mu_B$ moment for the vacancy (Figure~\ref{fig:figure2}, Table~\ref{tab:table1}).

Similarly, the corresponding electronic structure depends on the exchange-correlation functional. While a metallic solution emerges within the PBE and HSE06 functionals, the SCAN functional correctly predicts a gapped band structure.  This result is consistent with the diverging resistivity observed in irradiation induced defected graphene with about 10$^{12}$ cm$^{-2}$ vacancy concentration, which indicate insulating behaviour.~\citep{PhysRevLett.102.236805} This was latter theoretically corroborated in graphene with high vacancy concentrations and an insulating solution with a large bandgap was proposed.~\citep{PhysRevB.86.075402,PhysRevLett.114.246801}  Further, while the defect induced V$_{\sigma}$ state shows a large spin-splitting above 2 eV originating from the JT reconstruction, the splitting for the V$_{\pi}$ is small, about 250 meV (Figure~\ref{fig:figure2}). This prediction is consistent with the recent STM experiment, where the spin-polarized $\pi$ density of states (DOS) peaks at the defect site are observed to be separated by 20-60 meV.~\citep{PhysRevLett.117.166801} It should be noted here that in such experiments the V$_\pi$ spin-splitting may be affected by a metastable vacancy geometry or by its interaction with the substrate.

\begin{table}[!b]
\caption{Magnetic moment of the defect complexes in Figure~\ref{fig:figure1} severely depends on the treatment of exchange-correlation energy. We compare the moments calculated with conventional PBE, hybrid HSE06 ($a_{_{\rm HF}}$=0.25) and SCAN meta-GGA functionals, where the later better describes the moment. While the HSE06 calculations were carried out with 4$\times$3  ($\sqrt{3}$, 3)$a_0$ supercell, the 10$\times$5  ($\sqrt{3}$, 3)$a_0$ supercell was used for the calculations with PBE and SCAN density functionals.}
\label{tab:table1}
\begin{tabular}{L{5cm}C{0.9cm}C{0.9cm}C{0.9cm}}
\hline
\hline
\multirow{2}{*}{Defect} & \multicolumn{3}{c}{Magnetic moment ($\mu_B$)} \\
                        				& PBE & HSE & SCAN \\
\hline
(a) Monovacancy                            & 1.5   & 1.6    & 2.0 \\
(b) N adatom                               & 0.9   & 1.0    & 1.0 \\
(c) Graphitic N                            & 0.0   & 0.1    & 1.0 \\
(d) Meta-graphitic N pair (N$_{\rm AA}$)   & 0.0   & 0.4    & 2.0 \\
(e) Para-graphitic N pair (N$_{\rm AB'}$)  & 0.0   & 0.0    & 0.0 \\
(f) Graphitic 3N (G3N)                     & 0.9   & 1.0    & 1.0  \\
(g) Triazinic  N (T)                       & 0.9   & 1.0    & 1.0 \\
(h) Trimerized pyridinic N (TPY)           & 0.2   & 1.0    & 1.0 \\
(i) Trimerized pyrolic N (TPI)             & 1.0   & 1.0    & 1.0  \\
\hline
\hline
\end{tabular}
\end{table}

\begin{figure}[!t]
 \begin{center}
\rotatebox{0}{\includegraphics[width=0.46\textwidth]{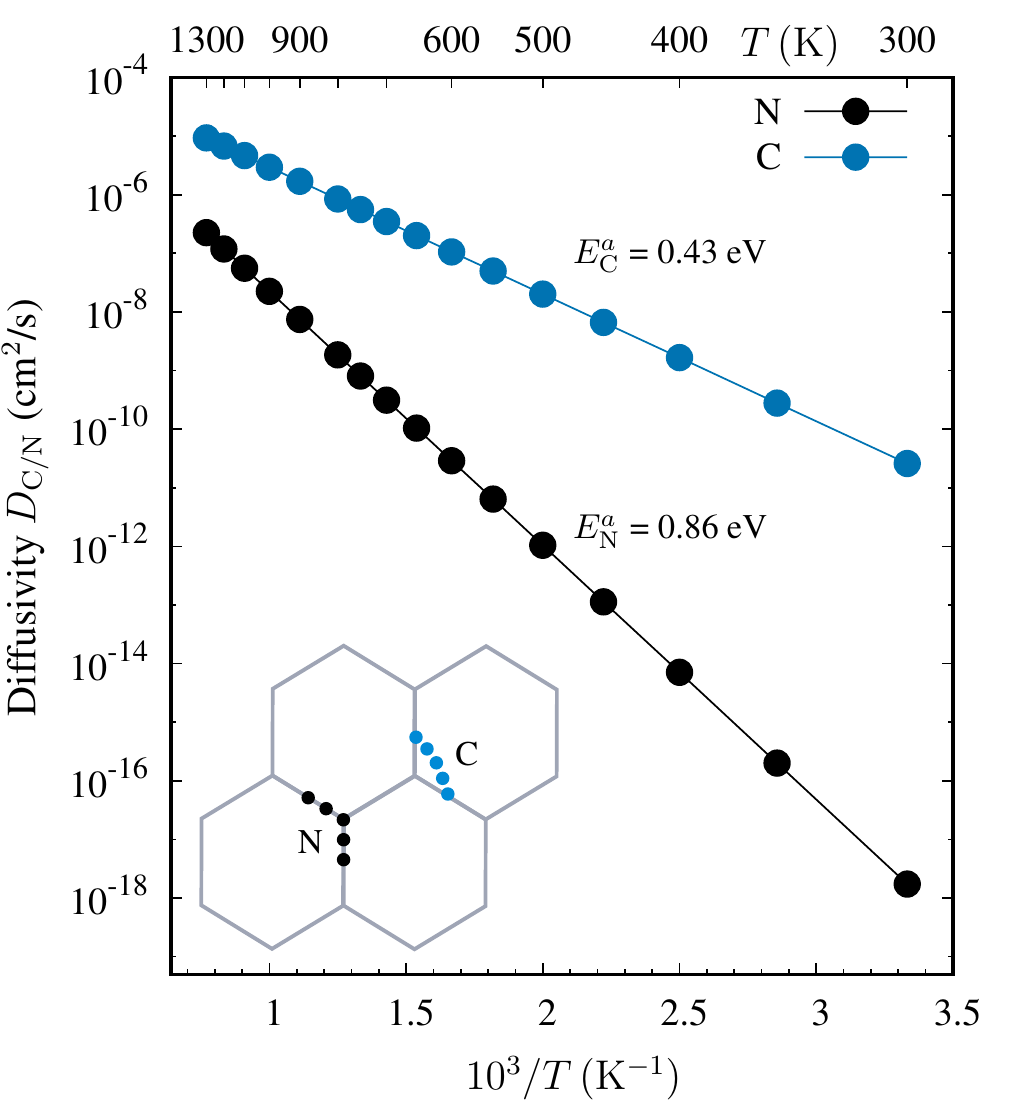}}
 \caption{Adatom diffusivity $D_{\rm C/N}$ calculated within the harmonic approximation, $D_{\rm X} = \frac{1}{4}a'_{_{\rm X}} \Gamma_0 e^{-E^a_{_{\rm X}}/k_BT}$, where $a'_{_{\rm X}}$ is the jump distance for the $\rm X=C/N$ diffusion, and $E^a_{_{\rm X}}$ is the corresponding activation barrier. $\Gamma_0$ $(= \Pi_i \nu^{\rm I}_i / \Pi_j \nu^{\rm TS}_j)$ is the jump prefactor, where $\nu^{\rm I}_i$ and $\nu^{\rm TS}_j$ are the frequencies of harmonic vibrational modes corresponding to the structures at the initial and saddle point. Very high diffusivity of C and N adatoms on graphene, especially in the temperature range 800-1300 K relevant to growth and annealing conditions,~\citep{jacs.6b12934,ja512897m,jacs.6b12934,Liu2016,LI2015460,Miao2016,C3NR34291C,Zhao999,nl2031037,cs200652y,srep00586,ja408463g,10.1021/nn506074u} indicates self-healing of the vacancy defects and promotes the formation of graphitic N-complexes. Both adatoms are absorbed at the bridge-site, and the two different diffusion mechanisms are shown in the inset.
}
 \label{fig:figure3}
 \end{center}
 \end{figure}

We find the vacancy magnetism to be robust against carrier doping with a realistic concentration,~\citep{nnano.2008.67} as the position of the Fermi level is tuned (Figure~\ref{fig:figure2}). The graphene with a neutral V$_1(5|9)$ defect is intrinsically hole doped, and one extra electron on a large 10$\times$5  ($\sqrt{3}$, 3)$a_0$ supercell corresponds to 1.89$\times$10$^{13}$ cm$^{-2}$ carrier density. The extra electron occupies the unfilled V$_{\pi}^{\downarrow}$ state destroying the $\pi$-electron imbalance in the two bipartite sublattices. Thus,  the $\pi$-component of the vacancy moment is completely quenched (Figure~\ref{fig:figure2}). In contrast, the $\sigma$-component of the magnetic moment is protected against carrier doping. Increasing the electron doping beyond 1.89$\times$10$^{13}$ cm$^{-2}$ does not further affect the magnetic moment (Figure~\ref{fig:figure2}) At higher electron density the unoccupied bulk $\pi$-states get filled, while the unoccupied V$_{\sigma}^{\downarrow}$ state lies much higher in energy and is impossible to populate with any realistic carrier density. The carrier concentration dependent differential charge density corroborates this picture.~\citep{PhysRevB.98.075439} The scenario with hole doping is different only qualitatively. The $\sigma$-component is unaffected, and hole doping does not quench the $\pi$-moment completely indicating the particle-hole asymmetry in defected graphene  (Figure~\ref{fig:figure2}). The semi-localized V$_{\pi}^{\uparrow}$ state is only partially depopulated and a fractional $\pi$-moment survives.~\citep{ncomms3010,PhysRevB.90.245420} Therefore, while the $\pi$ magnetism in graphene vacancy could be completely quenched with sufficient carrier doping, the spin-$\frac{1}{2}$ $\sigma$-magnetism is robust.

In the context of differential defect abundance in N-doped graphene, it is relevant to investigate the C and N adatom diffusion on graphene lattice. The C and N adatoms are absorbed at the bridge site with 1.51 and 0.9 eV binding energy. We calculate the activation barrier for adatom diffusion within climbing image nudged elastic band method.~\citep{1.1329672} In agreement with the previous calculations,~\citep{PhysRevLett.91.017202, jp512886t} the calculated activation barriers are found to be 0.43 and 0.86 eV, respectively, for C and N diffusion (Supplemental Material).~\citep{supple} Such low activation barrier results into a very high diffusivity $D_{\rm C/N}$ at the experimental growth and annealing temperatures (Figure~\ref{fig:figure3}), and the adatoms will quickly recombine with the graphene vacancies. Therefore, due to high adatom mobility, it is expected that the graphene vacancies will be annihilated to decrease in concentration and from graphitic N-complexes, while annealed at high-temperature or non-equilibrium growth conditions.~\citep{jacs.6b12934,ja512897m,jacs.6b12934,Liu2016,LI2015460,Miao2016,C3NR34291C,Zhao999,nl2031037,cs200652y,srep00586,ja408463g,10.1021/nn506074u} This prediction is consistent with the experimental observation of pyridinic to graphitic defect conversion during postannealing at high-temperature.~\citep{nl2031037}  Further, the diffusion mediated self-healing has been observed in scanning transmission electron microscopy.~\citep{nl300985q,nn401113r}

{\em N-adatom.} The chemisorbed N-adatom [Figure~\ref{fig:figure1}(b)] is a p-type defect with the impurity state above the Fermi level. A localized magnetic moment of 1 $\mu_B$ is generated from the in-plane N-orbitals (Table~\ref{tab:table1}). However, due to the high N diffusion on graphene (Figure~\ref{fig:figure3}), such adatom defect may not be present and effectively contribute to magnetism in graphene samples that are prepared and annealed at high-temperature. In contrast, high N diffusion will further promote the formation of graphitic N-complexes. 

{\em Graphitic nitrogen defects.} The appearance of Dirac point (DP, $E_{\rm D}$) in graphene band structure is very sensitive to external perturbations, and the breaking of sublattice symmetry induces a substantial gap at the DP.~\citep{Ohta951,PhysRevB.76.073103,nmat2154b} Thus, all the graphitic N-defects are expected to generate a gap at $E_{\rm D}$. Single graphitic nitrogen [Figure~\ref{fig:figure1}(c)] introduces an impurity state near the Fermi level and induces a 0.60 eV gap at $E_{\rm D}$. An overall semiconducting electronic structure emerges and the  results are consistent with the STM measurement.~\citep{nn506074u} Further, an exchange splitting of 18 meV is observed between the majority and minority spin channels, which results in itinerant $\pi$-magnetism along with a small 0.06 $\mu_B$ moment localized at the N-site. However, the total moment generated by this defect strongly depends on the treatment of exchange-correlation energy (Table~\ref{tab:table1}) and defect concentration shown in Supplemental Material.~\citep{supple}  While the SCAN functional correctly generates 1 $\mu_B$ moment, the moment strongly depends on the fractional exact exchange $a_{_{\rm HF}}$ in HSE06 functional. While the conventional HSE06 functional ($a_{_{\rm HF}}$=0.25) underestimates the moment (Table~\ref{tab:table1}), it gradually increases with increasing $a_{_{\rm HF}}$ and generates 1 $\mu_B$ moment for $a_{_{\rm HF}}$=1.  In contrast, in agreement with an earlier report,~\citep{PhysRevB.87.085441}  the conventional PBE functional fails to produce any spin-splitting and a non-magnetic solution emerges.  Therefore, the SCAN meta-GGA functional provides the correct magnetic picture over other exchange-correlational functionals.


The graphitic complexes with two nitrogens N$_{\rm AA}$ and N$_{\rm AB'}$ have been experimentally identified [Figure~\ref{fig:figure1}(d) and (e)].~\citep{cm102666r,srep00586,ja408463g,10.1021/nn506074u} Due to the sublattice asymmetry in these n-type complexes, the DP is destroyed, and within the SCAN functional a 0.31 and 0.50 eV gap emerges at the $E_{\rm D}$ for N$_{\rm AA}$ and N$_{\rm AB'}$ complexes, respectively.  The overall electronic structure is found to be defect specific as a half-metallic (semiconducting) solution is observed for the N$_{\rm AA}$  (N$_{\rm AB'}$) defect in the large 10$\times$5  ($\sqrt{3}$, 3)$a_0$ supercell. The overall results here are in agreement with the experimental results as the character of electronic DOS near the Fermi level is consistent with the STS measurements.~\citep{srep00586,10.1021/nn506074u} 
The exchange-split bands generate itinerant magnetism and a small moment of 0.06--0.08 $\mu_B$ is generated at the N-sites, which are coupled ferromagnetically (antiferromagnetically) while the nitrogen atoms occupy the same (different) sublattice. This fact results into 2 and 0 $\mu_B$ moments for the N$_{\rm AA}$ and N$_{\rm AB'}$ pairs, respectively (Table~\ref{tab:table1}). While the results from the different exchange-correlation functionals are compared, a similar picture emerges as discussed for the single graphitic N-defect. 


\begin{figure}[!t]
 \begin{center}
\rotatebox{0}{\includegraphics[width=0.48\textwidth]{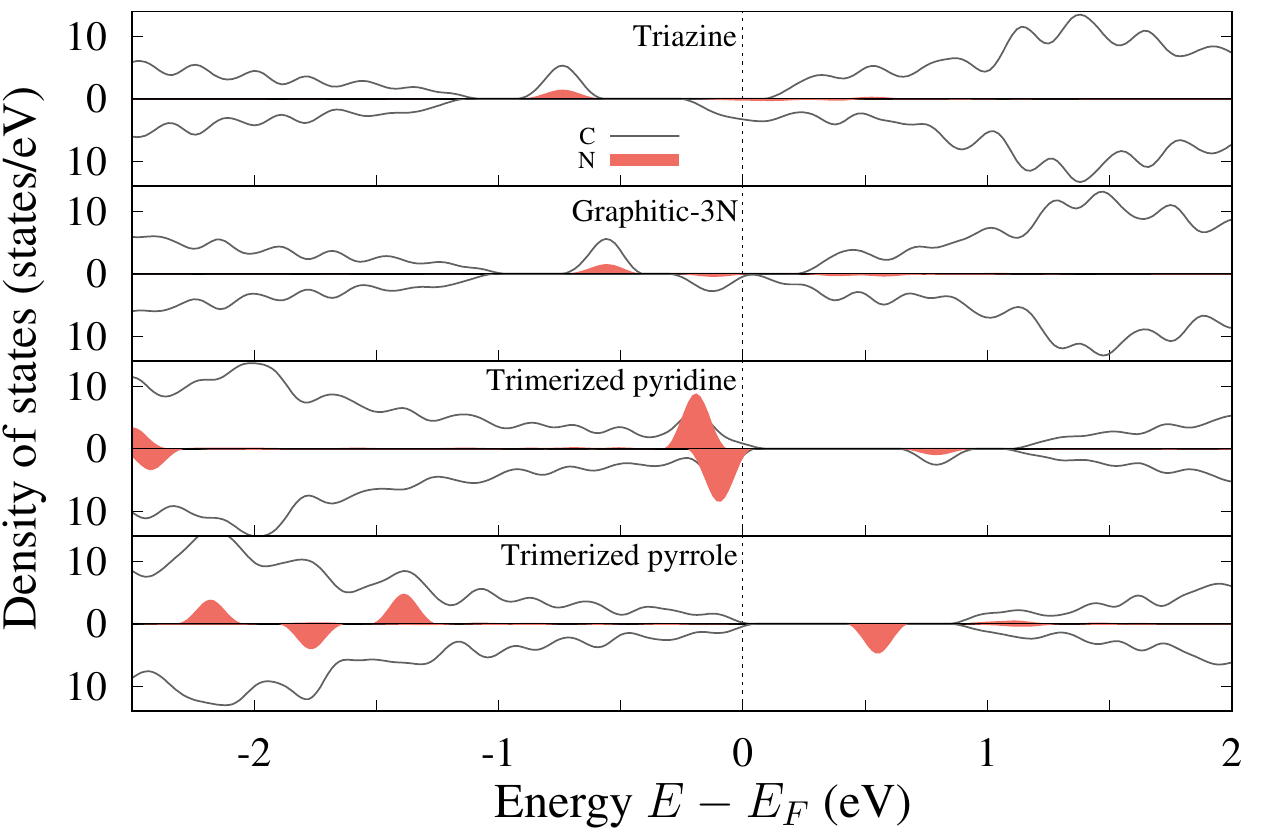}}
 \caption{Spin-polarized projected density of states calculated using the (10$\times$5)-($\sqrt{3}$,3)$a_0$ supercell for the graphitic and vacancy containing N-complexes. Substitutional N-doping opens up a gap at the Dirac point $E_{\rm D}$ and the electronic structure becomes n-type. In contrast, the vacancy containing defects display p-type semiconducting behaviour. 
}
 \label{fig:figureDOS}
 \end{center}
 \end{figure}

\begin{figure}[!t]
\begin{center}
\rotatebox{0}{\includegraphics[width=0.46\textwidth]{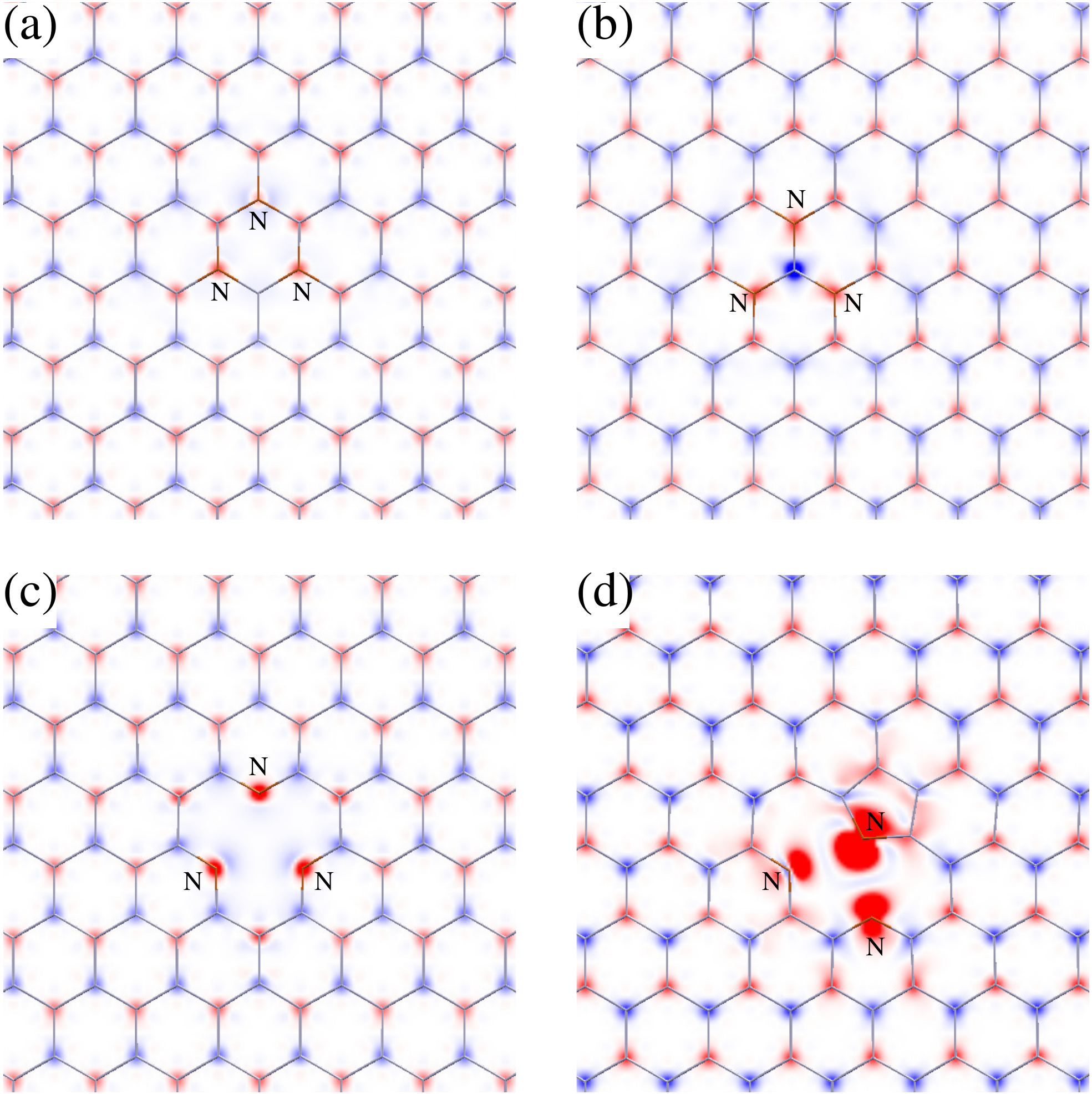}}
\caption{Magnetization density for graphitic and vacancy containing N-complexes: (a) triazine, (b) graphitic-3N, (c) trimerized pyridine and (d) trimerized pyrrolic defect. Red and blue colour indicate the up and down spin density, respectively. While the magnetism in graphitic N-complexes is mostly itinerant in nature, the magnetization density demonstrates that a very significant fraction of the moment is localized at the N-sites for the vacancy containing N-complexes.
}
\label{fig:figureM}
\end{center}
\end{figure}

While the triazine (T) complex [Figure~\ref{fig:figure1}(g)] is thermodynamically slightly more favorable over the graphitic-3N (G3N) complex [Figure~\ref{fig:figure1}(g)], both of these defects are formed at the non-equilibrium growth conditions.\citep{supple} The graphitic-3N complex could also be produced when a highly mobile carbon adatom diffuses into the lattice vacancy in the trimerized pyridinic and pyrrolic  complexes in the bulk regions of graphene. In both the defects [Figure~\ref{fig:figure1}(f) and Figure~\ref{fig:figure1}(g)], all three nitrogen atoms occupy the same sublattice leading to a sublattice asymmetry, which in turn destroys the DP and consequently open up a gap at the $E_{\rm D}$ (Figure~\ref{fig:figureDOS}).  The overall electronic structure is half-metallic for the T-complex, and the same is found to be semiconducting for the G3N-complex  (Figure~\ref{fig:figureDOS}). It is important to note that such half-metallicity induced asymmetric electron transport for different spin components could have significant implication in spin-based devices. To gain further insight into the magnetism, we investigate the differential charge density and the corresponding magnetization density (Figure~\ref{fig:figureM}). The C$\rightarrow$N $\pi$-electron transfer creates Lieb's imbalance and an exchange split $\pi$-band emerges, which is responsible for the itinerant magnetism that is confirmed form the magnetization density [Figure~\ref{fig:figureM}(a) and (b)]. In addition, a hybridized nitrogen impurity state is observed near the Fermi level, which is responsible for a tiny N-moment of $\sim$ 0.05 $\mu_B$. Note the C-atom at the centre of G3N-defect acquire a comparatively larger localized moment of 0.25 $\mu_B$ [Figure~\ref{fig:figureM}(b)]. These together result into a total moment of 1 $\mu_B$  for these graphitic defect complexes (Table~\ref{tab:table1}). 

 
Thus, in general, the electronic and magnetic structure of the different graphitic defect complexes are remarkably similar. First, all of them open up a gap at the $E_{\rm D}$ due to sublattice asymmetry, which is consistent with the the angle-resolved photoemission spectroscopy results.~~\citep{nl2031037} Second, the exchange-split $\pi$-band generates itinerant magnetism, while the localized moment at the N-site is small about 0.03--0.09 $\mu_B$.  While the overall electronic structure of the graphitic complexes are defect specific, all of them are found to be n-type, which are in overall agreement with the STS results.~\citep{srep00586,10.1021/nn506074u, nn506074u}

{\em Vacancy containing nitrogen defects.} Now we explore the defect complexes wherein the N atoms are incorporated in a vacancy. The trimerized pyridine (TPY), where three N atoms occupy the same graphene sublattice surrounding a vacancy [Figure~\ref{fig:figure1}(h)] is energetically favoured over the mono and dimerized pyridines.~\citep{PhysRevB.84.245446} The 1.33 \AA\ C$-$N bonds are consistent with experimental values for pyridinic complexes. The N atoms form an equilateral triangle with 2.61 \AA\ sides, which is larger than the triangular unrelaxed vacancy with 2.46 \AA. Further, this complex is found to be p-type and the Bader analysis indicates a 0.9-1.12 $e$/N charge transfer to nitrogen from carbon. 

The magnetism in such vacancy containing nitrogen defects is rather intricate, and a detailed investigation into the electronic DOS (Figure~\ref{fig:figureDOS}) and  the magnetization density [Figure~\ref{fig:figureM}(c)] unravel the origin of 1 $\mu_B$ moment for this complex. Unlike the bare vacancy, the $\sigma$ component to the moment is completely ruled out due to the presence of occupied $\sigma$-state that originates from the undercoordinated N atoms (Figure~\ref{fig:figureDOS}). In contrast, the moment is of entirely $\pi$ character and originates from the localized N-$p_z$ and itinerant C-$p_z$ states with an exchange-splitting of 11 meV.  As a result, a sizeable localized moment of 0.15 $\mu_B$ is observed at the N-sites for this complex and a fraction of the total moment originates form the itinerant magnetism [Figure~\ref{fig:figureM}(c)]. In contrast to the SCAN functional, the PBE results provide a very contrasting picture. First, the moment is substantially underestimated (Table~\ref{tab:table1}), and further, the moment originates from the localized $\sigma$-state at the N-sites without any $\pi$ contributions.  However, the STM and STS investigations of the monomerized and trimerized pyridine reveal an electron density distribution that is consistent with the $p_z$ states of the vacancy defect and thus corroborate the results obtained with the SCAN functional.~\citep{nn506074u,PhysRevB.86.035436,PhysRevLett.117.166801} 

The trimerized pyrrolic (TPI) complex is formed when a single N atom is absorbed in a divacancy to form a pentagon and rest of the undercoordinated C atoms are also substituted by N in Figure~\ref{fig:figure1}(i). The pentagonal C$-$N bonds (1.41 \AA) are larger in comparison with the other C$-$N bonds (1.33 \AA). Similar to the TPY-complex and bare vacancy, this complex has p-type character. The electronic DOS calculated within the PBE and SCAN functionals show qualitatively similar features, and the unoccupied $\sigma$-states originating from the N-impurities appear just above the Fermi level (Figure~\ref{fig:figureDOS}). The calculated magnetization density indicates that a large fraction $\sim$ 80\% of the total 1 $\mu_B$ moment is localized on the nitrogen atoms and only a small fraction originates from the delocalized $\pi$-band [Figure~\ref{fig:figureM}(d)]. Further, due to the proximity of the unoccupied $\sigma$-states to the Fermi level, a slight perturbation through the charge doping will adversely affect this localized $\sigma$ component of the moment. 

Therefore, the magnetism in the vacancy containing N-complexes is fundamentally different from the graphitic defects. While the magnetism in graphitic defects is entirely itinerant, a very significant fraction of the moment is localized at the N-sites in vacancy containing N-complexes.  Thus, in the absence of a strong exchange path, one would not expect that such localized magnetic moments in these vacancy containing defect complexes to contribute significantly to the high-temperature ferromagnetism. Indeed, the role of this particular defect on the observed ferromagnetism has remained controversial. ~\citep{ja512897m,jacs.6b12934,LI2015460}

\subsection{Magnetic interactions}
Although nitrogen-doped graphene is reported to show ferromagnetism, the microscopic mechanism is not yet understood and resulted in conflicting arguments.~\citep{ja512897m,jacs.6b12934,Liu2016,LI2015460,Miao2016,C3NR34291C}  The appearance of a diverse Curie temperature with varied growth and annealing conditions makes the microscopic analysis very complicated as the differential defect abundance may fluctuate drastically in the samples. A long-range magnetic order in nitrogen doped graphene can only emerge when the moment generating defects are exchange coupled. Thus, we investigate the magnetic interactions between the abundant defect complexes.

The single vacancy in graphene lattice generates a robust semi-localized moment.  However, the interaction between the vacancy moments is sublattice dependent. Vacancies in the same sublattice are coupled ferromagnetically, whereas they prefer antiferromagnetic alignment when placed on different sublattices. Further, the magnetic interaction sharply falls to zero below 1 nm distance due to semi-localized nature of the moment.~\citep{SCOPEL20165} Thus, a randomly distributed vacancy in graphene will not lead to any magnetic ordering, which is consistent with the results predicting paramagnetism.~\citep{10.1038/nphys2183,PhysRevLett.105.207205,10.1063/1.3628245} Moreover, it should be noticed in this context that, the existence of a vacancy in the N-doped graphene is rare due to the very high C and N adatom mobility at the high-temperature growth conditions (Figure~\ref{fig:figure3}), which leads to self-healing of vacancy or the formation of graphitic N-defect.      

Single graphitic nitrogen defect is similar to the pristine vacancy. For two substitutional atoms on the (different) same sublattice, the spin-configuration is (anti-) ferromagnetic. Moreover, the interaction strength decays fast with increasing separation and follows an inverse power law, and the long-range interaction may develop through Ruderman-Kittel-Kasuya-Yosida (RKKY) mechanism.~\citep{PhysRevB.87.085441} Thus, a statistically random distribution of single graphitic nitrogen will not result in high-temperature ferromagnetism that has been reported so far.~\citep{ja512897m,jacs.6b12934,Liu2016,LI2015460,Miao2016,C3NR34291C}

\begin{figure}[!t]
 \begin{center}
\rotatebox{0}{\includegraphics[width=0.48\textwidth]{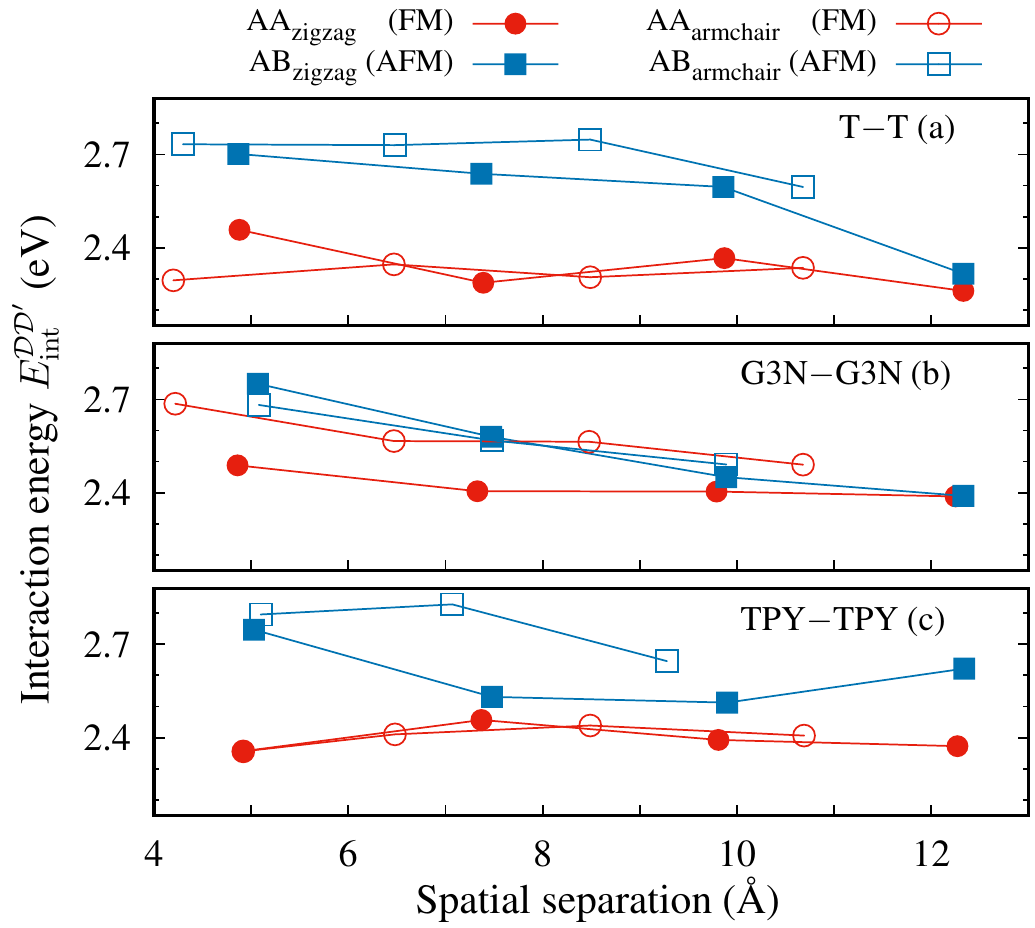}}
 \caption{Interaction energy between the N-complexes $E^{\mathcal{DD}'}_{\rm int}$ in graphene is calculated in the low concentration limit (3 at. \%) with varying distances and orientations along the zigzag and armchair directions.  Independent of the interacting defect types, the magnetic coupling is observed to be sublattice depended. The defects on the equivalent (opposite) sublattice are coupled ferromagnetically (antiferromagnetically). Such sublattice dependent magnetic interaction will not lead to the observed ferromagnetism in nitrogen-doped graphene. In contrast, these complexes interact very differently in the high N-concentration limit, which is discussed in the text. 
}
 \label{fig:figure_int}
 \end{center}
 \end{figure}

A very complex picture emerges when we explore the interactions between the larger defects. Here, we address two distinct regimes of N-doping that is motivated by several experimental observations.~\citep{jacs.6b12934, Liu2016,Miao2016,unpublished} It has been unanimously pointed out in the literature that ferromagnetism emerges at a critical N-concentration around 5 at.~\% and above which the saturation magnetization and coercive field increase further with increasing the concentration. Thus, we consider N-concentration below 5 at.~\% and in the 8$-$13 at.~\% range, and investigate how the concentration influences the magnetic interaction. In this regard, we calculate the interaction energy $E^{\mathcal{DD}'}_{\rm int}$ between two magnetic defects $\mathcal{D}$ and $\mathcal{D}'$   as, $E^{\mathcal{DD}'}_{\rm int} = (E^{\mathcal{DD}'}_{\rm m} -  E^{\mathcal{DD}'}_{\rm nm}) - (E^{\mathcal{D}}_{\rm m} -  E^{\mathcal{D}}_{\rm nm}) - (E^{\mathcal{D}'}_{\rm m} -  E^{\mathcal{D}'}_{\rm nm})$ and shown in the Figure~\ref{fig:figure_int}. The energies corresponding to the magnetic and non-magnetic solutions for the interacting defects are $E^{\mathcal{DD}'}_{\rm m}$ and $E^{\mathcal{DD}'}_{\rm nm}$, while $E^{\mathcal{D/D}'}_{\rm m}$ and $E^{\mathcal{D/D}'}_{\rm nm}$ represent magnetic and non-magnetic energies of the isolated defects.    

\begin{figure*}[!t]
 \begin{center}
\rotatebox{0}{\includegraphics[width=0.96\textwidth]{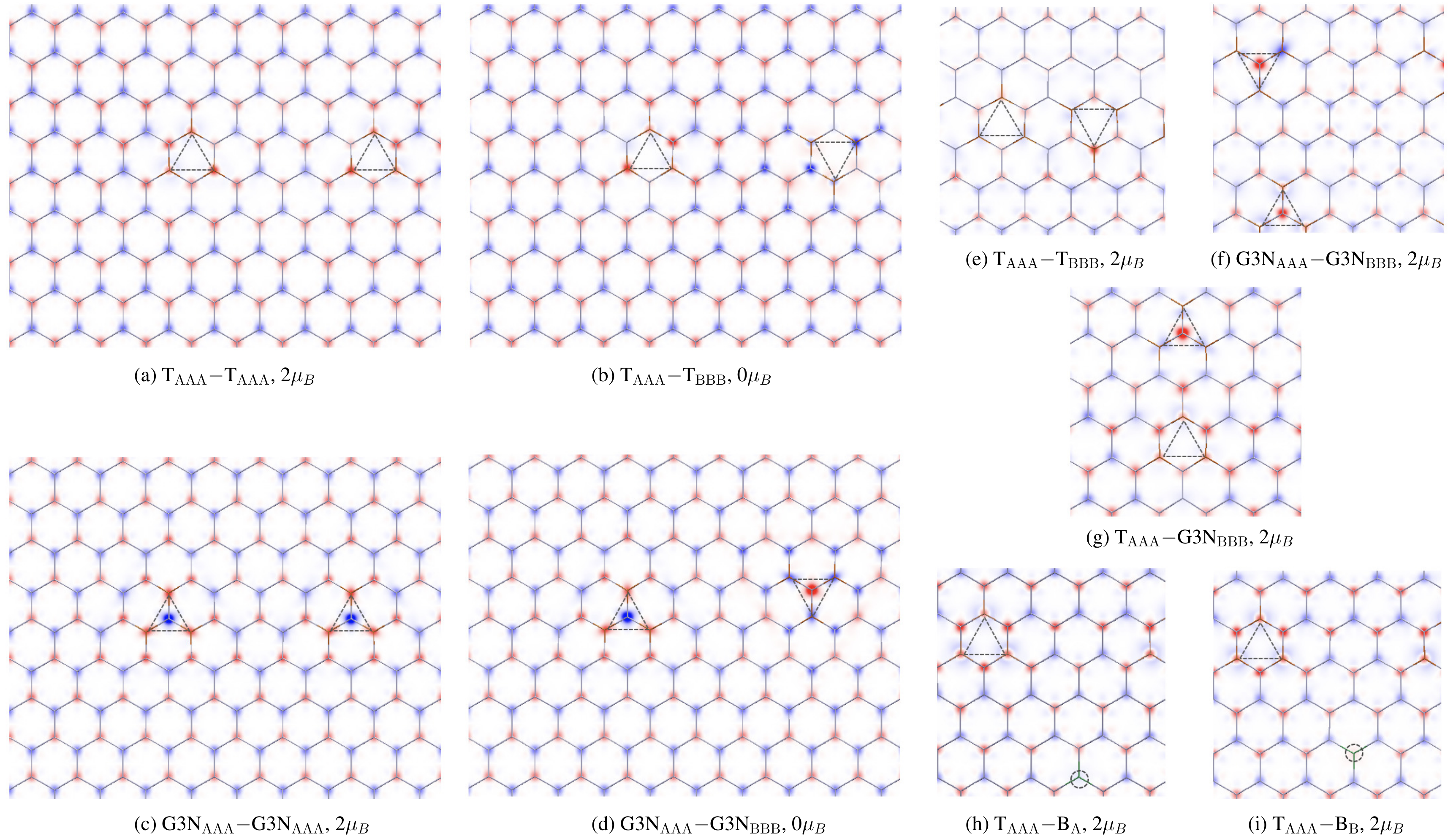}}
 \caption{The calculated magnetization density reveals the corresponding magnetic interactions between the defects, which are highlighted with a triangle or circle. In the low N-concentration limit of 3 at. \%, the graphitic defects follow the usual sublattice dependent magnetic interaction. (a)--(d) The defects on the (opposite) same sublattice are (anti-) ferromagnetically coupled. However, the picture changes at the high N-concentration limit (12.5 at. \%) and crucial exceptions to the usual sublattice dependent magnetic occur. (e)--(g) In specific crystallographic direction of graphene lattice, the defects are coupled ferromagnetically, regardless of their sublattice. Such ferromagnetic interaction between the graphitic N-defects is responsible for the observed ferromagnetism. (h)--(i) The interaction between the T-complex and the single graphitic B-defect is always ferromagnetic, independent of the sublattice. Thus, B co-doping promotes ferromagnetism in N-doped graphene and increase the saturation magnetization, which is consistent with the recent experimental observation.~\cite{unpublished}  Red and blue colour indicate the up and down spin density, respectively.  
}
 \label{fig:figureI}
 \end{center}
 \end{figure*}

First, we discuss the long-range magnetic interaction between the graphitic defects in the low N-concentration limit and investigate how it is affected by the spatial separation and direction between the defects (Figure~\ref{fig:figure_int}). Two interacting T$-$T, G3N$-$G3N, and mixed T$-$G3N defects in a (10$\times$5)-($\sqrt{3}$,3)$a_0$ supercell represents 3 at.~\% N-concentration. At such a low concentration and for all these types of defect interactions, we notice some general trends [Figure~\ref{fig:figure_int}(a) and~\ref{fig:figure_int}(b)].  (i) These defects mostly prefer to occupy the same graphene sublattice, (ii) owing to the itinerant nature of the individual moments in these graphitic defects, the interaction between them is long-range, and (iii) the nature of magnetic interaction between the defects is entirely dictated by their respective sublattice [Figure~\ref{fig:figureI}(a)--(d)]. For the defects on the equivalent (opposite) graphene sublattice, the magnetic interaction is found to be ferromagnetic FM (antiferromagnetic AFM). The picture remains unaltered regardless of the spatial separation and relative orientation between the defects along the armchair and zigzag directions (Figure~\ref{fig:figure_int}). Although the nature of the magnetic moment in these defects is completely different, the sublattice dependent magnetic interaction is in complete agreement with results predicted by the RKKY interaction between the localized moments on the bipartite graphene lattice.~\citep{PhysRevB.76.184430, PhysRevLett.99.116802} While these defects mostly prefer to occupy the same sublattice $\mathcal{D}_{\rm A}\mathcal{D}'_{\rm A}$ over the $\mathcal{D}_{\rm A}\mathcal{D}'_{\rm B}$ configuration, during the non-equilibrium growth process the defects are equivalently distributed over the two graphene sublattice. Therefore, the sublattice dependent FM and AFM interaction will restrict the observation of FM order at such a low concentration of nitrogen. This picture is consistent with the experimental observation that FM ordering is not found below 5 at.~\% N-concentration.~\citep{jacs.6b12934}

Now we turn our attention onto high N-concentration, and two interacting T$-$T, G3N$-$G3N, and mixed T$-$G3N defects in a (4$\times$3)-($\sqrt{3}$,3)$a_0$ supercell correspond to 12.5 at.~\% N-doping. The sublattice dependent magnetism for such high N-concentration changes dramatically. In contrast to the AFM interaction, a FM interaction with spontaneous magnetization emerges while the defects that are placed on the different sublattice [Figure~\ref{fig:figureI}(e)--(g)].  However, such FM structure is found when the defects are interacting along a particular direction, which is along the zigzag direction for  T$-$T interaction [Figure~\ref{fig:figureI}(e)] and along the armchair direction for the G3N$-$G3N [Figure~\ref{fig:figureI}(f)] , and mixed T$-$G3N [Figure~\ref{fig:figureI}(g)] defect interactions. Thus, at high N-concentration, the graphitic T and G3N defects are responsible for the observed ferromagnetism. Further, the concentration of the graphitic T-complex may be increased by using the triazine-based precursor during the growth using the chemical vapour deposition technique.~\citep{10.1021/nn402102y} Moreover, a suitable postannealing strategy may convert the vacancy containing pyridinic and pyrrolic defects to graphitic G3N-complex.~\citep{nl2031037} Therefore, we conclude that such sample preparation strategies may give rise to robust ferromagnetism with improved Curie temperature.  This discussion is consistent with the recent observation of room-temperature ferromagnetism in turbostratic N-doped graphene films, where the T-defect is found to be the most abundant complex.~\citep{unpublished}

Further, in the high concentration limit of 8.3 at.~\% N-doping, we investigate the interactions between the two N$_{\rm XX}$ defects, and between the T-complex and a single graphitic N-defect. Both these interactions follow the usual sublattice dependent magnetic couplings. While the N$_{\rm AA}$$-$ N$_{\rm AA}$ and T$_{\rm AAA}$$-$N$_{\rm A}$ interactions are found to be ferromagnetic, the N$_{\rm AA}$$-$ N$_{\rm BB}$ and T$_{\rm AAA}$$-$N$_{\rm B}$ are observed to be antiferromagnetic. Therefore, such defect-defect interactions are not responsible for the observed ferromagnetism in the N-doped graphene.
   
In contrast to the graphitic N-defects, the vacancy containing complexes do not contribute to ferromagnetism. Moreover, in accordance with the experimental observation,~\citep{nl2031037} we have discussed earlier that the C-adatom is highly mobile on the graphene lattice (Figure~\ref{fig:figure3}), and thus the vacancy containing TPY-defect complex will be converted into the graphitic G3N-complex during high-temperature growth and annealing. Therefore, the concentration of this defect will be negligibly small. However, for completeness, we have investigated the magnetic interaction between two TPY-complexes in both the low and high concentration limits. Unlike the graphitic T- and G3N-complexes and independent of N-concentration, the TPY$-$TPY magnetic coupling is AFM (FM) for defects on the opposite (equivalent) sublattice. Further, the picture is independent of defect orientation and separation [Figure~\ref{fig:figure_int}(c)]. Thus, this vacancy containing defect will not play any role in the ferromagnetism in N-doped graphene.       

\subsection{Effects of boron co-doping}
Motivated by the recent observation of room-temperature ferromagnetism in boron and nitrogen doped turbostratic carbon films,~\cite{unpublished} we further investigate the effect of B co-doping on the ferromagnetism. Notably, we consider a small B co-doping in the limit of high N-concentration that is responsible for the observed ferromagnetism. Due to very-low B-concentration, we have only considered graphitic single B-defect and ignored the possibility of the larger B-complexes. We also ignored the chemisorbed B-defect, as in this configuration, the B-adatom is expected to be very mobile similar to the C and N adatoms (Figure~\ref{fig:figure3}), and consequently form single graphitic B-defect.

In contrast to the single N-defect, the graphitic B-defect leads to p-type behaviour due to electron deficiency. The C--B bonds are found to elongated (1.49 \AA) compared to the C--N bonds (1.41 \AA) in the graphitic N-defect and a B $\rightarrow$ C charge transfer takes place. However, similar to the single graphitic N-defect, the B-defect also results in an itinerant magnetism with 1 $\mu_B$ moment, where a very small 0.03 $\mu_B$ moment is localized at the B-site. In contrast to the repulsive and sublattice dependent N$-$N magnetic interactions, the N$-$B interaction is found to be attractive, and a completely compensated AFM structure emerges, which is independent of the sublattice.  Further, we investigate the interaction between this graphitic B-defect and the prevalent graphitic N-defects, T and G3N complexes with 1 $\mu_B$ moment each. Surprisingly, the B$_{\rm A/B}$ affects the overall picture significantly. The interaction between these graphitic defects and the B$_{\rm A/B}$ is found to be attractive, which is opposite to their interaction with single graphitic N-defect. Independent of their respective sublattice, the interacting T$_{\rm AAA}$$-$B$_{\rm A/B}$ [Figure~\ref{fig:figureI}(h)--(i)] and G3N$_{\rm AAA}$$-$B$_{\rm A/B}$ configurations generate a total moment of 2 $\mu_B$, while for all cases, the compensated AFM magnetic structure is found to be  energetically much unfavourable. A closer look at the magnetization density reveals that the resultant 2 $\mu_B$ moment arises due to the sublattice imbalance of the itinerant C-$p_z$ electrons. This finding is consistent with the recent experimental observation of increased saturation magnetization in N-doped graphene due to B co-doping.~\cite{unpublished} Therefore, we argue that small B co-doping will enhance ferromagnetism in nitrogen-doped graphene.   

\section{Summary}
The microscopic origin of ferromagnetism that is observed in N-doped graphene has remained elusive.~\citep{ja512897m,jacs.6b12934,Liu2016,LI2015460,Miao2016,C3NR34291C} The difficulty to comprehend arises due to the complex dependence of magnetism on N-content and on the corresponding differential defect distribution originating from the varied growth and annealing details. In this context, we systematically investigate the moment formation and the intricate magnetic interaction between various defect complexes within the first-principles calculations. The C and N adatom diffusion on graphene plays a very important role in healing vacancy containing defects and generating graphitic N-complexes. The magnetism in the graphitic N-complexes are found to be fundamentally different from the vacancy containing complexes. While an itinerant magnetism emerges in the graphitic N-complexes, a very significant fraction of the moment is localized at the N-sites for the later. By comparing the results with PBE and HSE06 exchange-correlation functional and available experimental data, we conclude that the SCAN semilocal meta-GGA density functional provides a better description of the electronic structure and magnetism in defected graphene. We further demonstrate that the vacancy magnetism is robust against carrier doping. 

The magnetic interaction between the defects are found to be intricate and complex, which depends on the defect type, N-concentration, defect orientation and the corresponding distance between them. Regardless of the defect type and at low N-concentration, the magnetic interaction shows a generic sublattice dependence independent of defect orientation and distance. In agreement with the RKKY interaction leading to sublattice dependent magnetic interaction in bipartite graphene lattice,~\citep{PhysRevB.76.184430, PhysRevLett.99.116802} the long-range coupling is found to be FM (AFM) for the defects on the equivalent (opposite) sublattice. Thus, at low N-content, the statistical distribution of defects on both graphene sublattice will not produce FM ordering. This result is consistent with experimental observation.~\citep{ja512897m} Interestingly, the picture drastically changes at the high N-concentration owing to a direction dependent and sublattice independent FM interaction with spontaneous magnetization for the interaction between the graphitic T and G3N defects. While we reject the other type of defects through a comprehensive investigation, we conclude that graphitic T and G3N defects are predominantly responsible for the observed ferromagnetism. Further, we predict that an experimental strategy to increase the concentration of these graphitic defects will produce robust ferromagnetism with improved Curie temperature. Finally, we demonstrate that a  small amount of boron co-doping will further enhance the ferromagnetism in N-doped graphene. Thus, while the present results help us to understand the scattered and often controversial experimental results, an overall microscopic understanding and a concurrent experimental synthesis strategy provided here will motivate new experiments to control magnetism in nitrogen and boron doped graphene.

\begin{acknowledgements}
We acknowledge stimulating discussions with S. B. Ogale and R. Mandal.  M.K. acknowledges funding from the Science and Engineering Research Board through Ramanujan Fellowship and the Department of Science and Technology through Nano Mission project SR/NM/TP-13/2016. The supercomputing facilities at the Centre for Development of Advanced Computing, Pune; Inter University Accelerator Centre, Delhi; and at the Center for Computational Materials Science, Institute of Materials Research, Tohoku University are greatly acknowledged.
\end{acknowledgements}


%

\end{document}